\pdfoutput=1
\documentclass[12pt,a4paper]{article}

\usepackage{ifthen} 
\newboolean{pdflatex}
\setboolean{pdflatex}{true} 
\usepackage{booktabs}

\newboolean{articletitles}
\setboolean{articletitles}{true} 

\newboolean{uprightparticles}
\setboolean{uprightparticles}{false} 

\def\paperauthors{LHCb collaboration} 
\def\paperasciititle{Measurement of CP asymmetries in B0b to DsmDp and Bsb to Dsp Dm decays} 
\def\papertitle{Measurement of $\CP$ asymmetries in $\Bdb \to \Dsm \Dp$ and $\Bsb \to \Dsp \Dm$ decays}
\def\paperkeywords{{High Energy Physics}, {LHCb}} 
\def\papercopyright{\the\year\ CERN for the benefit of the LHCb collaboration} 
\def\paperlicence{CC BY 4.0 licence}
\def\paperlicenceurl{https://creativecommons.org/licenses/by/4.0/}

\newif\ifEnableSectionTOCLinks
\EnableSectionTOCLinksfalse

\usepackage[top=1in, bottom=1.25in, left=1in, right=1in]{geometry}

\usepackage{microtype}
\usepackage{lineno}  
\usepackage{xspace} 
\usepackage{caption} 

\usepackage{graphicx}  
\usepackage{color}
\usepackage{colortbl}
\graphicspath{{./figs/}} 

\usepackage{amsmath} 
\usepackage{amssymb}
\usepackage{amsfonts}
\usepackage{upgreek} 

\newcommand*\patchAmsMathEnvironmentForLineno[1]{%
\expandafter\let\csname old#1\expandafter\endcsname\csname #1\endcsname
\expandafter\let\csname oldend#1\expandafter\endcsname\csname
end#1\endcsname
 \renewenvironment{#1}%
   {\linenomath\csname old#1\endcsname}%
   {\csname oldend#1\endcsname\endlinenomath}%
}
\newcommand*\patchBothAmsMathEnvironmentsForLineno[1]{%
  \patchAmsMathEnvironmentForLineno{#1}%
  \patchAmsMathEnvironmentForLineno{#1*}%
}
\AtBeginDocument{%
\patchBothAmsMathEnvironmentsForLineno{equation}%
\patchBothAmsMathEnvironmentsForLineno{align}%
\patchBothAmsMathEnvironmentsForLineno{flalign}%
\patchBothAmsMathEnvironmentsForLineno{alignat}%
\patchBothAmsMathEnvironmentsForLineno{gather}%
\patchBothAmsMathEnvironmentsForLineno{multline}%
\patchBothAmsMathEnvironmentsForLineno{eqnarray}%
}

\usepackage[pdftex,
            pdfauthor={\paperauthors},
            pdftitle={\paperasciititle},
            pdfkeywords={\paperkeywords}]{hyperref}
\usepackage{hyperxmp}
\hypersetup{
    pdfcopyright={Copyright (C) \papercopyright},
    pdflicenseurl={\paperlicenceurl}
}

\usepackage[bottom,flushmargin,hang,multiple]{footmisc}

\usepackage[all]{hypcap} 

\usepackage{xspace}
\usepackage{upgreek}

\def\lhcb   {\mbox{LHCb}\xspace}

\def\MagUp {\mbox{\em Mag\kern -0.05em Up}\xspace}

\ifthenelse{\boolean{uprightparticles}}%
{

 \def\Pmu         {\ensuremath{\upmu}\xspace}
 \def\Pnu         {\ensuremath{\upnu}\xspace}
 
 \def\Ppi         {\ensuremath{\uppi}\xspace}

 \def\Ppsi        {\ensuremath{\uppsi}\xspace}

 \def\PDelta      {\ensuremath{\Delta}\xspace}
 \def\PXi         {\ensuremath{\Xi}\xspace}
 \def\PLambda     {\ensuremath{\Lambda}\xspace}
 \def\PSigma      {\ensuremath{\Sigma}\xspace}
 \def\POmega      {\ensuremath{\Omega}\xspace}
 \def\PUpsilon    {\ensuremath{\Upsilon}\xspace}
 \let\oldPi\Pi
 \def\PPi         {\ensuremath{\oldPi}\xspace}

 \def\PB      {\ensuremath{\mathrm{B}}\xspace}
 \def\PD      {\ensuremath{\mathrm{D}}\xspace}
 \def\PJ      {\ensuremath{\mathrm{J}}\xspace}
 \def\PK      {\ensuremath{\mathrm{K}}\xspace}
 \def\Pb      {\ensuremath{\mathrm{b}}\xspace}
 \def\Pc      {\ensuremath{\mathrm{c}}\xspace}
 \def\Pd      {\ensuremath{\mathrm{d}}\xspace}

 \def\Pp      {\ensuremath{\mathrm{p}}\xspace}

 \def\Ps      {\ensuremath{\mathrm{s}}\xspace}
 \def\Pt      {\ensuremath{\mathrm{t}}\xspace}
 \def\Pu      {\ensuremath{\mathrm{u}}\xspace}
 \def\thebaroffset{0.0em}
}
{

 \def\Pmu         {\ensuremath{\mu}\xspace}
 \def\Pnu         {\ensuremath{\nu}\xspace}
 
 \def\Ppi         {\ensuremath{\pi}\xspace}

 \def\Ppsi        {\ensuremath{\psi}\xspace}
 
 \mathchardef\PDelta="7101
 \mathchardef\PXi="7104
 \mathchardef\PLambda="7103
 \mathchardef\PSigma="7106
 \mathchardef\POmega="710A
 \mathchardef\PUpsilon="7107
 \mathchardef\PPi="7105
 \def\PB      {\ensuremath{B}\xspace}
 \def\PD      {\ensuremath{D}\xspace}
 \def\PJ      {\ensuremath{J}\xspace}
 \def\PK      {\ensuremath{K}\xspace}
 \def\Pb      {\ensuremath{b}\xspace}
 \def\Pc      {\ensuremath{c}\xspace}
 \def\Pd      {\ensuremath{d}\xspace}

 \def\Pp      {\ensuremath{p}\xspace}

 \def\Ps      {\ensuremath{s}\xspace}
 \def\Pt      {\ensuremath{t}\xspace}
 \def\Pu      {\ensuremath{u}\xspace}
 \def\thebaroffset{0.18em}
}
\newcommand{\offsetoverline}[2][\thebaroffset]{\kern #1\overline{\kern -#1 #2}}%

\makeatletter
\ifcase \@ptsize \relax
  \newcommand{\miniscule}{\@setfontsize\miniscule{4}{5}}
\or
  \newcommand{\miniscule}{\@setfontsize\miniscule{5}{6}}
\or
  \newcommand{\miniscule}{\@setfontsize\miniscule{5}{6}}
\fi
\makeatother

\DeclareRobustCommand{\optbar}[1]{\shortstack{{\miniscule (\rule[.5ex]{1.25em}{.18mm})}
  \\ [-.7ex] $#1$}}

\def\mun        {{\ensuremath{\Pmu^-}}\xspace} 

\def\neub       {{\ensuremath{\overline{\Pnu}}}\xspace}

\def\neumb      {{\ensuremath{\neub_\mu}}\xspace}

\def\uquark    {{\ensuremath{\Pu}}\xspace}

\def\dquark    {{\ensuremath{\Pd}}\xspace}
\def\dquarkbar {{\ensuremath{\overline \dquark}}\xspace}

\def\squark    {{\ensuremath{\Ps}}\xspace}
\def\squarkbar {{\ensuremath{\overline \squark}}\xspace}

\def\cquark    {{\ensuremath{\Pc}}\xspace}
\def\cquarkbar {{\ensuremath{\overline \cquark}}\xspace}

\def\bquark    {{\ensuremath{\Pb}}\xspace}

\def\tquark    {{\ensuremath{\Pt}}\xspace}

\def\pion   {{\ensuremath{\Ppi}}\xspace}
\def\piz    {{\ensuremath{\pion^0}}\xspace}
\def\pip    {{\ensuremath{\pion^+}}\xspace}
\def\pim    {{\ensuremath{\pion^-}}\xspace}

\def\kaon    {{\ensuremath{\PK}}\xspace}

\def\KorKbar {\kern \thebaroffset\optbar{\kern -\thebaroffset \PK}{}\xspace}

\def\Kp      {{\ensuremath{\kaon^+}}\xspace}
\def\Km      {{\ensuremath{\kaon^-}}\xspace}

\def\KS      {{\ensuremath{\kaon^0_{\mathrm{S}}}}\xspace}

\def\Dbar    {{\ensuremath{\offsetoverline{\PD}}}\xspace}
\def\D       {{\ensuremath{\PD}}\xspace}

\def\DorDbar {\kern \thebaroffset\optbar{\kern -\thebaroffset \PD}\xspace}
\def\Dz      {{\ensuremath{\D^0}}\xspace}
\def\Dzb     {{\ensuremath{\Dbar{}^0}}\xspace}
\def\Dp      {{\ensuremath{\D^+}}\xspace}
\def\Dm      {{\ensuremath{\D^-}}\xspace}
\def\Dpm     {{\ensuremath{\D^\pm}}\xspace}
\def\Dmp     {{\ensuremath{\D^\mp}}\xspace}
\def\DpDm    {\ensuremath{\Dp {\kern -0.16em \Dm}}\xspace}

\def\Dstarp  {{\ensuremath{\D^{*+}}}\xspace}

\def\Dsp     {{\ensuremath{\D^+_\squark}}\xspace}
\def\Dsm     {{\ensuremath{\D^-_\squark}}\xspace}
\def\Dspm    {{\ensuremath{\D^{\pm}_\squark}}\xspace}
\def\Dsmp    {{\ensuremath{\D^{\mp}_\squark}}\xspace}
\def\Dss     {{\ensuremath{\D^{*+}_\squark}}\xspace}
\def\Dssp    {{\ensuremath{\D^{*+}_\squark}}\xspace}

\def\Dsspm   {{\ensuremath{\D^{*\pm}_\squark}}\xspace}

\def\B       {{\ensuremath{\PB}}\xspace}
\def\Bbar    {{\ensuremath{\offsetoverline{\PB}}}\xspace}

\def\BorBbar {\kern \thebaroffset\optbar{\kern -\thebaroffset \PB}\xspace}
\def\Bz      {{\ensuremath{\B^0}}\xspace}
\def\Bzb     {{\ensuremath{\Bbar{}^0}}\xspace}
\def\Bd      {{\ensuremath{\B^0}}\xspace}
\def\Bdb     {{\ensuremath{\Bbar{}^0}}\xspace}
\def\BdorBdbar {\kern \thebaroffset\optbar{\kern -\thebaroffset \Bd}\xspace}

\def\Bub     {{\ensuremath{\B^-}}\xspace}

\def\Bm      {{\ensuremath{\Bub}}\xspace}

\def\Bs      {{\ensuremath{\B^0_\squark}}\xspace}
\def\Bsb     {{\ensuremath{\Bbar{}^0_\squark}}\xspace}
\def\BsorBsbar {\kern \thebaroffset\optbar{\kern -\thebaroffset \Bs}\xspace}

\def\Bds     {{\ensuremath{\B_{(\squark)}^0}}\xspace}
\def\Bdsb    {{\ensuremath{\Bbar{}_{(\squark)}^0}}\xspace}
\def\BdorBs  {\Bds}
\def\BdorBsbar  {\Bdsb}

\def\BdorBsorbar {\optbar{\B}{}^0_{(\squark)} \xspace}

\def\jpsi     {{\ensuremath{{\PJ\mskip -3mu/\mskip -2mu\Ppsi}}}\xspace}

\def\Y#1S{\ensuremath{\PUpsilon{(#1S)}}\xspace}

\def\proton      {{\ensuremath{\Pp}}\xspace}

\def\LorLbar     {\kern \thebaroffset\optbar{\kern -\thebaroffset \PLambda}\xspace}

\newcommand{\decay}[2]{\mbox{\ensuremath{#1\!\to #2}}\xspace}

\def\to                 {\ensuremath{\rightarrow}\xspace}

\def\grpsuthree {{\ensuremath{\mathrm{SU}(3)}}\xspace}

\def\CP                {{\ensuremath{C\!P}}\xspace}

\def\Vcd  {{\ensuremath{V_{\cquark\dquark}^{\phantom{\ast}}}}\xspace}
\def\Vtd  {{\ensuremath{V_{\tquark\dquark}^{\phantom{\ast}}}}\xspace}

\def\Vcs  {{\ensuremath{V_{\cquark\squark}^{\phantom{\ast}}}}\xspace}
\def\Vts  {{\ensuremath{V_{\tquark\squark}^{\phantom{\ast}}}}\xspace}

\def\Vcbs  {{\ensuremath{V_{\cquark\bquark}^\ast}}\xspace}
\def\Vtbs  {{\ensuremath{V_{\tquark\bquark}^\ast}}\xspace}

\def\AT#1     {\ensuremath{A_{\mathrm{T}}^{#1}}\xspace}           

\def\C#1      {\ensuremath{\mathcal{C}_{#1}}\xspace}                       
\def\Cp#1     {\ensuremath{\mathcal{C}_{#1}^{'}}\xspace}                    
\def\Ceff#1   {\ensuremath{\mathcal{C}_{#1}^{\mathrm{(eff)}}}\xspace}        
\def\Cpeff#1  {\ensuremath{\mathcal{C}_{#1}^{'\mathrm{(eff)}}}\xspace}       
\def\Ope#1    {\ensuremath{\mathcal{O}_{#1}}\xspace}                       
\def\Opep#1   {\ensuremath{\mathcal{O}_{#1}^{'}}\xspace}                    

\newcommand{\nospaceunit}[1]{\ensuremath{\text{#1}}}
\newcommand{\aunit}[1]{\ensuremath{\text{\,#1}}}

\newcommand{\tev}{\aunit{Te\kern -0.1em V}\xspace}
\newcommand{\gev}{\aunit{Ge\kern -0.1em V}\xspace}
\newcommand{\mev}{\aunit{Me\kern -0.1em V}\xspace}
\newcommand{\kev}{\aunit{ke\kern -0.1em V}\xspace}
\newcommand{\ev}{\aunit{e\kern -0.1em V}\xspace}

\newcommand{\mevc}{\ensuremath{\aunit{Me\kern -0.1em V\!/}c}\xspace}
\newcommand{\gevc}{\ensuremath{\aunit{Ge\kern -0.1em V\!/}c}\xspace}
\newcommand{\mevcc}{\ensuremath{\aunit{Me\kern -0.1em V\!/}c^2}\xspace}
\newcommand{\gevcc}{\ensuremath{\aunit{Ge\kern -0.1em V\!/}c^2}\xspace}

\def\mum  {\ensuremath{\,\upmu\nospaceunit{m}}\xspace}

\def\fb   {\ensuremath{\aunit{fb}}\xspace}
\def\invfb   {\ensuremath{\fb^{-1}}\xspace}

\def\gsim{{~\raise.15em\hbox{$>$}\kern-.85em
          \lower.35em\hbox{$\sim$}~}\xspace}
\def\lsim{{~\raise.15em\hbox{$<$}\kern-.85em
          \lower.35em\hbox{$\sim$}~}\xspace}

\def\sqs   {\ensuremath{\protect\sqrt{s}}\xspace}

\def\pt         {\ensuremath{p_{\mathrm{T}}}\xspace}

\def\ptot       {\ensuremath{p}\xspace}
\def\et         {\ensuremath{E_{\mathrm{T}}}\xspace}

\def\evtgen     {\mbox{\textsc{EvtGen}}\xspace}

\def\geant      {\mbox{\textsc{Geant4}}\xspace}

\def\photos     {\mbox{\textsc{Photos}}\xspace}

\def\pythia     {\mbox{\textsc{Pythia}}\xspace}

\def\tell1  {TELL1\xspace}
\def\ukl1   {UKL1\xspace}

\newcommand{\lhcborcid}[1]{\href{https://orcid.org/#1}{\hspace*{0.1em}\raisebox{-0.45ex}{\includegraphics[width=1em]{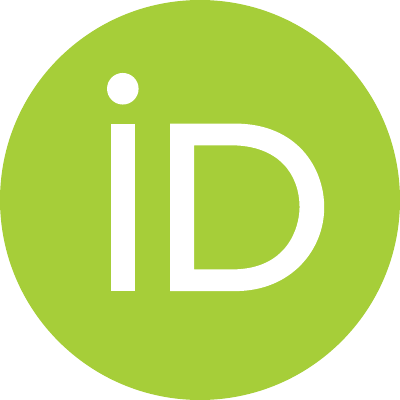}}}}

\hypersetup{
  colorlinks   = true, 
  urlcolor     = blue, 
  linkcolor    = blue, 
  citecolor    = red   
}

\ifEnableSectionTOCLinks
    \usepackage[explicit]{titlesec} 
    
    \let\oldcontentsline\contentsline
    \renewcommand

    \titleformat{\section}{\normalfont\Large\bf}{\hyperlink{tocsection.\thesection}{{\thesection} \parbox[t]{\dimexpr\textwidth-1pc}{#1}}}{1pc}{}

    \titleformat{\subsection}{\normalfont\bf}{\hyperlink{tocsubsection.\thesubsection}{{\thesubsection} \parbox[t]{\dimexpr\textwidth-1pc}{#1}}}{1pc}{}

    \titleformat{name=\section,numberless}[display]{}{}{0pt}{\normalfont\Huge\bfseries #1}
\fi

\usepackage{cite} 
\usepackage{mciteplus}

\usepackage[compat=1.1.0]{tikz-feynman}
\usepackage{contour}
\usepackage{tikz}
\usetikzlibrary{angles,quotes,arrows.meta,quotes}
\usepackage{makecell}
\usepackage{subfig}
\usepackage{longtable} 

\begin{document}

\renewcommand{\thefootnote}{\fnsymbol{footnote}}
\setcounter{footnote}{1}


\begin{titlepage}
\pagenumbering{roman}

\vspace*{-1.5cm}
\centerline{\large EUROPEAN ORGANIZATION FOR NUCLEAR RESEARCH (CERN)}
\vspace*{1.5cm}
\noindent
\begin{tabular*}{\linewidth}{lc@{\extracolsep{\fill}}r@{\extracolsep{0pt}}}
\ifthenelse{\boolean{pdflatex}}
{\vspace*{-1.5cm}\mbox{\!\!\!\includegraphics[width=.14\textwidth]{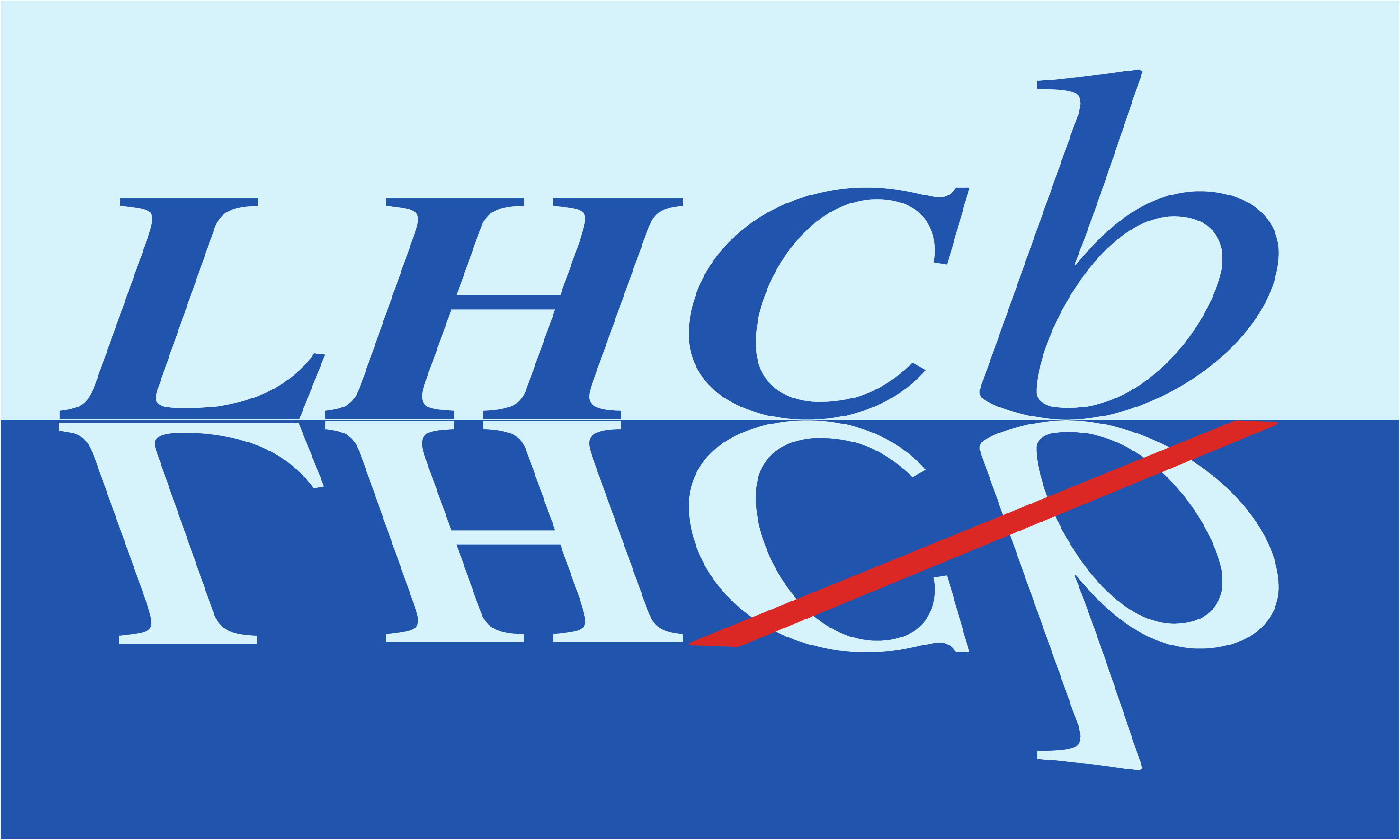}} & &}%
{\vspace*{-1.2cm}\mbox{\!\!\!\includegraphics[width=.12\textwidth]{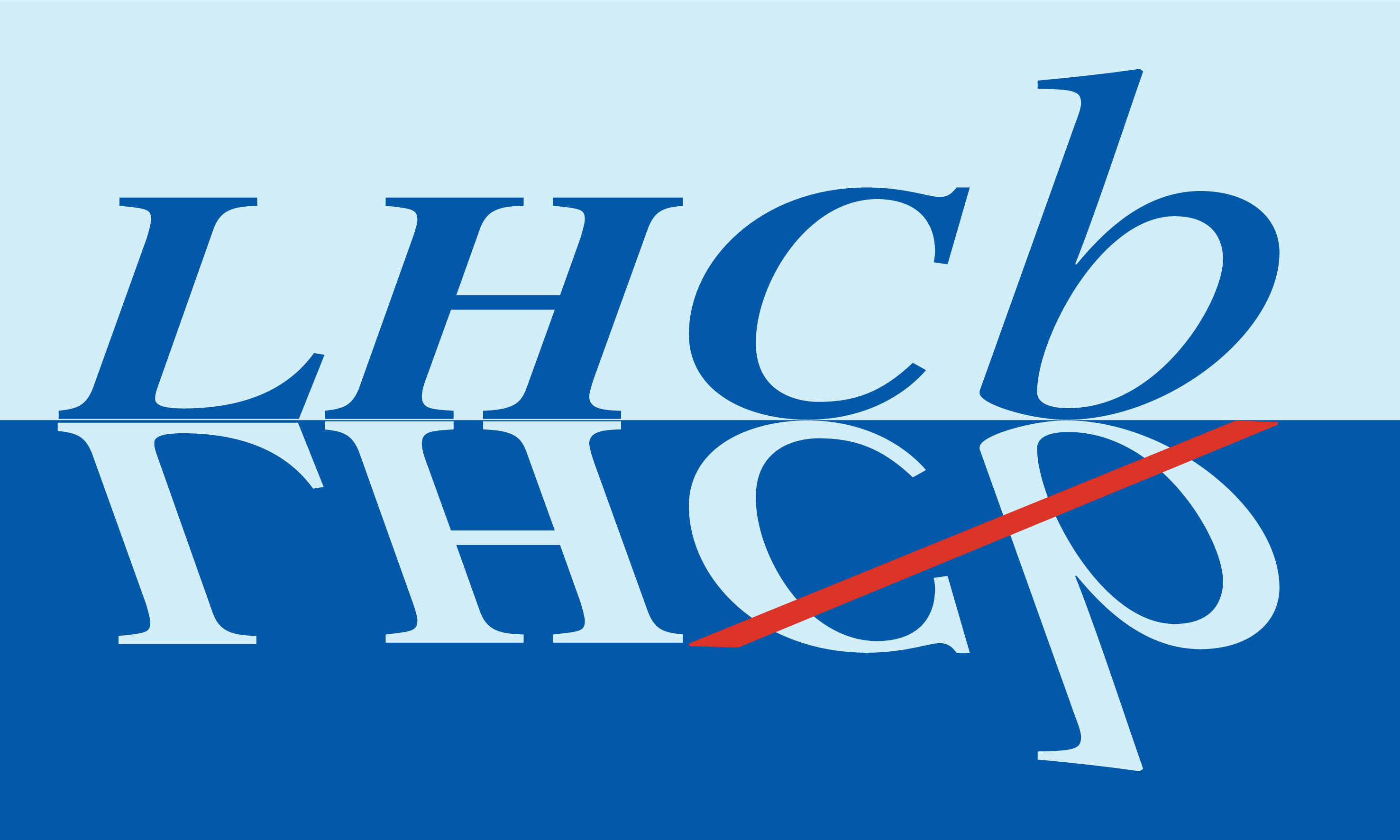}} & &}%
\\
 & & CERN-EP-2026-015 \\  
 & & LHCb-PAPER-2025-037 \\  
 & & 30 March 2026 \\ 
 & & \\
\end{tabular*}

\vspace*{4.0cm}

{\normalfont\bfseries\boldmath\huge
\begin{center}
  \papertitle 
\end{center}
}

\vspace*{2.0cm}

\begin{center}
\paperauthors\footnote{Authors are listed at the end of this paper.}
\end{center}

\vspace{\fill}

\begin{abstract}
  \noindent
  Measurements of the combined $\CP$ asymmetries in $\Bdb \to \Dsm \Dp$ and $\Bsb \to \Dsp \Dm$ decays are made using proton-proton collision data collected by the LHCb experiment, corresponding to an integrated luminosity of $9\invfb$. 
  The measurements are found to be
\begin{align*}
\begin{split}
A_{\CP}(\Bdb\to \Dsm\Dp) &= 0.0009 \pm 0.0053 \pm 0.0040, \\
A_{\CP}(\Bsb\rightarrow \Dsp \Dm ) &= 0.103\phantom{0} \pm 0.053\phantom{0} \pm 0.010,
\end{split}
\end{align*}
 where the first and second uncertainties are statistical and systematic, respectively. 
This is the first measurement of this asymmetry in \Bsb decays, and the most precise measurement to date for \Bzb decays. Both measurements are found to be consistent with $\CP$ symmetry. 
  
\end{abstract}

\vspace*{2.0cm}

\begin{center}
  Submitted to JHEP 
\end{center}

\vspace{\fill}

{\footnotesize 
\centerline{\copyright~\papercopyright. \href{\paperlicenceurl}{\paperlicence}.}}
\vspace*{2mm}

\end{titlepage}


\newpage
\setcounter{page}{2}
\mbox{~}
%
%
%
%

\renewcommand{\thefootnote}{\arabic{footnote}}
\setcounter{footnote}{0}

\cleardoublepage

\pagestyle{plain} 
\setcounter{page}{1}
\pagenumbering{arabic}

\section{Introduction}
\label{sec:Introduction}

Studies of $\CP$ violation in the hadronic decays of $B$ mesons are a vital probe of the Cabibbo--Kobayashi--Maskawa (CKM) paradigm within the Standard Model (SM), and moreover a powerful tool in the search for beyond Standard Model (BSM) physics. 
Of particular importance are decays mediated by $b\to \cquark\cquarkbar\dquark$ and $b\to \cquark\cquarkbar\squark$ transitions, which are crucial to the measurement of $\beta_{(s)}$, the phase of the combination of CKM matrix elements $\Vcd\Vcbs/\Vtd \Vtbs$ and $\Vts\Vtbs/\Vcs\Vcbs$ for $\beta$ and $\beta_s$, respectively, the precise determination of which requires control over subleading loop-level contributions to the decay amplitudes~\cite{fleischer1999extracting, jung2015standard}. 

A promising avenue for understanding higher-order contributions to $b\to \cquark \cquarkbar\dquark$ and \mbox{$b\to \cquark \cquarkbar\squark$} transitions is the study of $\CP$ violation in $\decay{B}{D\Dbar}$ decays. 
Within this programme, time-dependent studies of decays such as $\Bs\to\Dsp\Dsm$ play an analogous role to $\Bs\to \jpsi\phi$ in determining the angle $\beta_s$~\cite{Bel2015, jung2015standard}, while time-integrated measurements of $\CP$ asymmetries in $\Bzb\to\Dsm\Dp$ and $\Bsb\to\Dsp\Dm$ decays are used to control the loop-level contributions, the leading and subleading quark-level diagrams which are shown in Fig.~\ref{B0bar_feynman}. 
Furthermore, a global analysis of \CP asymmetries and branching fractions in $\decay{B_{(s)}}{D^{(*)}_{(s)}\Dbar{}^{(*)}_{(s)}}$ decays \cite{davies2023bar}, exploiting the approximate symmetries between the flavours of light quarks ($\grpsuthree_F$) in the relevant amplitudes, may be used in searches for possible BSM  contributions~\cite{Fu_2011, Li_2010, Lu_2011, PhysRevD.79.055004, xu2016studying}.

\begin{figure}[!b]
	\centering
	\[
    \resizebox{0.49\columnwidth}{!}{
    \begin{tikzpicture}
		\begin{feynman}
			\vertex (a1)                  {\bquark};
			\vertex[right=2cm of a1] (a2);
			\vertex[right=2cm of a2] (a3) {\cquark};
			\vertex[below=2em of a1] (b1) {\dquarkbar};
			\vertex[below=2em of a3] (b2) {\dquarkbar};
			\vertex[above=2em of a3] (c1) {\cquarkbar};
			\vertex[above=2em of c1] (c3) {\squark};
			\vertex[above=1em of c1] (ghost); 
            \vertex[left=1cm of ghost] (c2);
			\diagram* {
				(a1) -- [fermion] (a2) -- [fermion] (a3),
				(b2) -- [fermion] (b1),
				(c3) -- [anti fermion, out=180, in=45] (c2) -- [anti fermion, out=-45, in=180] (c1),
				(a2) -- [boson, bend left, edge label=\(W^{-}\)] (c2),
				
			};
			\draw [decoration={brace}, decorate] (b1.south west) -- (a1.north west)
			node [pos=0.5, left] {\(\Bzb \)};
			\draw [decoration={brace}, decorate] (c3.north east) -- (c1.south east)
			node [pos=0.5, right] {\(\Dsm\)};
			\draw [decoration={brace}, decorate] (a3.north east) -- (b2.south east)
			node [pos=0.5, right] {\(\Dp\)};
		\end{feynman}
	\end{tikzpicture}}
    \resizebox{0.49\columnwidth}{!}{
    \begin{tikzpicture}
		\begin{feynman}
			\vertex (a1) {\bquark};
			\vertex[right=1.333333333cm of a1](a2);
			\vertex[right=1.333333333cm of a2](a3);
            \vertex[right=1.333333333cm of a3](a4){\squark};

			\vertex[below=6em of a1] (b1) {\dquarkbar};
			\vertex[right=2cm of b1](b2);
			\vertex[right=2cm of b2](b3){\dquarkbar};
			\vertex[above=2em of b3] (c3) {\cquark};
            \vertex[above=1em of c3] (ghost) ;
            \vertex[left= 0.8cm of ghost] (c2); 
			\vertex[below=2em of a4] (c1) {\cquarkbar};
			\vertex at ($(a2)!0.5!(a3) + (0, -1cm)$) (c4);
			\diagram* {
				(a1) -- [fermion] (a2) -- [boson, edge label = \(W^{-}\)] (a3)-- [fermion] (a4),
				(b1) -- [anti fermion] (b3),
				(c1) -- [fermion, quarter right] (c2),
				(c2) -- [fermion, quarter right] (c3),
				(c4) -- [anti fermion, quarter left, edge label = \uquark{,}\cquark{,}\tquark] (a2),
				(a3) -- [anti fermion, quarter left] (c4),
				(c4) -- [boson, bend right, edge label' = $g$] (c2),
			};
			\draw [decoration={brace}, decorate] (b1.south west) -- (a1.north west)
			node [pos=0.5, left] {\(\Bdb\)};
			\draw [decoration={brace}, decorate] (a4.north east) -- (c1.south east)
			node [pos=0.5, right] {\(\Dsm\)};
			\draw [decoration={brace}, decorate] (c3.north east) -- (b3.south east)
			node [pos=0.5, right] {\(\Dp\)};
		\end{feynman}
	\end{tikzpicture}} 
	\]
	\[
    \resizebox{0.49\columnwidth}{!}{\begin{tikzpicture}
		\begin{feynman}
			\vertex (a1)                  {\bquark};
			\vertex[right=2cm of a1] (a2);
			\vertex[right=2cm of a2] (a3) {\cquark};
			\vertex[below=2em of a1] (b1) {\squarkbar};
			\vertex[below=2em of a3] (b2) {\squarkbar};
			\vertex[above=2em of a3] (c1) {\cquarkbar};
			\vertex[above=2em of c1] (c3) {\dquark};
			\vertex[above=1em of c1] (ghost); 
            \vertex[left=1cm of ghost] (c2);
			\diagram* {
				(a1) -- [fermion] (a2) -- [fermion] (a3),
				(b2) -- [fermion] (b1),
				(c3) -- [anti fermion, out=180, in=45] (c2) -- [anti fermion, out=-45, in=180] (c1),
				(a2) -- [boson, bend left, edge label=\(W^{-}\)] (c2),
				
			};
			\draw [decoration={brace}, decorate] (b1.south west) -- (a1.north west)
			node [pos=0.5, left] {\(\Bsb \)};
			\draw [decoration={brace}, decorate] (c3.north east) -- (c1.south east)
			node [pos=0.5, right] {\(\Dm\)};
			\draw [decoration={brace}, decorate] (a3.north east) -- (b2.south east)
			node [pos=0.5, right] {\(\Dsp\)};
		\end{feynman}
	\end{tikzpicture}}
    \resizebox{0.49\columnwidth}{!}{\begin{tikzpicture}
		\begin{feynman}
			\vertex (a1) {\bquark};
			\vertex[right=1.333333333cm of a1](a2);
			\vertex[right=1.333333333cm of a2](a3);
            \vertex[right=1.333333333cm of a3](a4){\dquark};

			\vertex[below=6em of a1] (b1) {\squarkbar};
			\vertex[right=2cm of b1](b2);
			\vertex[right=2cm of b2](b3){\squarkbar};
			\vertex[above=2em of b3] (c3) {\cquark};
            \vertex[above=1em of c3] (ghost) ;
            \vertex[left= 0.8cm of ghost] (c2); 
			\vertex[below=2em of a4] (c1) {\cquarkbar};
			\vertex at ($(a2)!0.5!(a3) + (0, -1cm)$) (c4);
			\diagram* {
				(a1) -- [fermion] (a2) -- [boson, edge label = \(W^{-}\)] (a3)-- [fermion] (a4),
				(b1) -- [anti fermion] (b3),
				(c1) -- [fermion, quarter right] (c2),
				(c2) -- [fermion, quarter right] (c3),
				(c4) -- [anti fermion, quarter left, edge label = \uquark{,}\cquark{,}\tquark] (a2),
				(a3) -- [anti fermion, quarter left] (c4),
				(c4) -- [boson, bend right, edge label' = $g$] (c2),
			};
			\draw [decoration={brace}, decorate] (b1.south west) -- (a1.north west)
			node [pos=0.5, left] {\(\Bsb\)};
			\draw [decoration={brace}, decorate] (a4.north east) -- (c1.south east)
			node [pos=0.5, right] {\(\Dm\)};
			\draw [decoration={brace}, decorate] (c3.north east) -- (b3.south east)
			node [pos=0.5, right] {\(\Dsp\)};
		\end{feynman}
	\end{tikzpicture}}
	\]
	\caption[$\BdorBsbar$ Feynman diagrams]{ (Left) Leading and (right) loop-level Feynman diagrams describing the weak (top) $\Bdb\to\Dsm\Dp$  and (bottom) $\Bsb\to\Dsp\Dm$ decays.}
	\label{B0bar_feynman}
\end{figure}

In this paper, $\CP$ violation in the flavour specific $\Bzb\to\Dsm\Dp$ and $\Bsb\to \Dsp\Dm$ decays is measured using the $\CP$ asymmetries 
\begin{align}
    \begin{split}
    \label{eq_acp_theor}
	A_{\CP}(\Bzb \rightarrow \Dsm\Dp) &\equiv \frac{\Gamma(\Bzb \rightarrow \Dsm\Dp) - \Gamma(\Bd \rightarrow \Dsp \Dm)}{\Gamma(\Bzb \rightarrow \Dsm\Dp) + \Gamma(\Bd \rightarrow \Dsp \Dm)}, \\ 
    A_{\CP}(\Bsb \rightarrow \Dsp \Dm ) &\equiv \frac{\Gamma(\Bsb \rightarrow \Dsp \Dm ) - \Gamma(\Bs \rightarrow \Dsm\Dp)}{\Gamma(\Bsb \rightarrow \Dsp \Dm ) + \Gamma(\Bs \rightarrow \Dsm\Dp)},
    \end{split}
\end{align}
where $\Gamma( \BdorBsorbar \to \Dspm \Dmp)$ are the time-integrated decay rates for each process. 
The most precise measurement to date of $A_{\CP}(\Bzb\to \Dsm\Dp)$ is from the Belle collaboration, and gives $A_{\CP}(\Bzb\to \Dsm\Dp) = -0.01\pm0.02$~\cite{rohrken2012measurements}, while $A_{\CP}(\Bsb\to \Dsp\Dm)$ is yet to be measured.
Driven by the relative CKM phases of the leading and subleading amplitudes, the $\CP$ asymmetry in $\Bz$ decay is expected to be very small, while for the $\Bsb$ decay the asymmetry could be as high as 18\%~\cite{jung2015standard,davies2023bar}.

The measurements presented in this paper are made using data corresponding to 1, 2 and $6\invfb$ of $pp$ collisions collected at centre-of-mass energies $\sqs = 7,8$ and $13\tev$, respectively. 
Both charm mesons are reconstructed in the charged three-body Cabibbo-favoured decays $\Dsm \to \Km\Kp\pim$ and $\Dp\to\Km\pip\pip$, which benefit from both a high reconstruction efficiency in LHCb and negligible $\CP$ violation in the SM. 
The $\CP$ asymmetries for both $\Bdb$ and $\Bsb$ decays are determined from the raw asymmetries, $A_\text{raw}$, which are defined as 
\begin{align}
    \begin{split}
	\label{A_raw_def}
        	A_{\text{raw}}(\Bdb\to \Dsm\Dp) &\equiv \frac{N(\Bdb\to \Dsm\Dp) - N(\Bd \to \Dsp\Dm)}{N(\Bdb \to \Dsm\Dp)+N(\Bd \to \Dsp\Dm)}, \\
            A_{\text{raw}}(\Bsb\to \Dsp\Dm) &\equiv \frac{N(\Bsb\to \Dsp\Dm) - N(\Bs\to \Dsm\Dp)}{N(\Bsb\to \Dsp\Dm)+N(\Bs\to \Dsm\Dp)},
    \end{split}
\end{align}
where $N$ is the measured yield of each decay. 
The measured asymmetries are affected by differences in detection efficiencies for particles of different charges, $A_\text{det}$, and differences in production cross-sections for different flavours of neutral $B$ mesons, $A_\text{prod}$, and can be expressed in terms of the asymmetries of interest, $A_{CP}$, and the nuisance asymmetries as 
\begin{align}
    \begin{split}
	\label{eq:cp_asymmetry}
    A_{\text{raw}}(\Bdb\to \Dsm\Dp) &= 
	A_{\CP}(\Bdb\to \Dsm\Dp) + A_{\text{det}}(\Bdb\to \Dsm\Dp) + A_{\text{prod}}(\Bdb),\\
	A_{\text{raw}}(\Bsb\to \Dsp\Dm) &= A_{\CP}(\Bsb\to \Dsp\Dm) + A_{\text{det}}(\Bsb\to \Dsp\Dm) + A_{\text{prod}}(\Bsb), 
    \end{split}
\end{align}
when the production and detection asymmetries are small, which here is a valid assumption. 
The detection asymmetry is given in terms of the charge-dependent efficiencies, $\varepsilon$, as 
\begin{align}
    \begin{split}
	\label{A_det_def}
	A_{\text{det}}(\Bdb) &\equiv \frac{\varepsilon(\Bdb\rightarrow \Dsm\Dp) - \varepsilon(\Bd\rightarrow \Dsp\Dm)}{\varepsilon(\Bdb\rightarrow \Dsm\Dp) + \varepsilon(\Bd\rightarrow \Dsp\Dm)}, \\
	A_{\text{det}}(\Bsb) &\equiv \frac{\varepsilon(\Bsb\rightarrow \Dsp\Dm) - \varepsilon(\Bs\rightarrow \Dsm\Dp)}{\varepsilon(\Bsb\rightarrow \Dsp\Dm) + \varepsilon(\Bs\rightarrow \Dsm\Dp)}.
    \end{split}
\end{align}
The detection asymmetries are determined using calibration data samples,  weighted to account for the differences in kinematics between the calibration and signal decays. 
The production asymmetries are defined as
\begin{equation}
	\label{A_Bs_prod_def}
	A_{\text{prod}}(\BdorBsbar) \equiv \frac{\sigma(\BdorBsbar) - \sigma(\BdorBs)}{\sigma(\BdorBsbar) + \sigma(\BdorBs)},
\end{equation}
where $\sigma(\BdorBsorbar)$ denotes the production cross-section of the relevant $B$ meson.

\section{Detector and simulation}
\label{sec:Detector}

The \lhcb detector~\cite{LHCb-DP-2008-001, LHCb-DP-2014-002} is a single-arm forward
spectrometer covering the \mbox{pseudorapidity} range $2<\eta <5$,
designed for the study of particles containing \bquark or \cquark
quarks. The detector, used to collect the data set for this analysis, includes a high-precision tracking system
consisting of a silicon-strip vertex detector surrounding the $pp$
interaction region~\cite{LHCb-DP-2014-001}, a large-area silicon-strip detector located
upstream of a dipole magnet with a bending power of about
$4{\mathrm{\,Tm}}$, and three stations of silicon-strip detectors and straw
drift tubes~\cite{LHCb-DP-2013-003,LHCb-DP-2017-001}
placed downstream of the magnet. The dipole generates a magnetic field that deflects oppositely charged particles in opposite directions and this can lead to detection asymmetries. Periodically
reversing the magnetic field polarity throughout the data-taking almost cancels
the effect. The configuration with the magnetic field pointing upwards (downwards), bends positively (negatively) charged particles
in the horizontal plane towards the centre of the LHC ring.

The tracking system provides a measurement of the momentum, \ptot, of charged particles with
a relative uncertainty that varies from 0.5\% at low momentum to 1.0\% at 200\gevc.
The minimum distance of a track to a primary $pp$ collision vertex (PV), the impact parameter (IP), 
is measured with a resolution of $(15+29/\pt)\mum$.
The \pt is the component of the momentum transverse to the beam, given in\,\gevc. 
Different types of charged hadrons are distinguished using information
from two ring-imaging Cherenkov detectors~\cite{LHCb-DP-2012-003}. 
Photons, electrons and hadrons are identified by a calorimeter system consisting of
scintillating-pad and preshower detectors, an electromagnetic
and a hadronic calorimeter. Muons are identified by a
system composed of alternating layers of iron and multiwire
proportional chambers~\cite{LHCb-DP-2012-002}.

The online collision event selection is performed by a trigger~\cite{LHCb-DP-2012-004}, 
which consists of a hardware stage, based on information from the calorimeter and muon
systems, followed by a software stage, which applies a full event
reconstruction.
At the hardware trigger stage, events are required to have a muon with high \pt or a hadron, photon or electron with high transverse energy (\et) in the calorimeters. 
The software trigger requires a two-, three- or four-track secondary vertex with a significant displacement from any primary $pp$ interaction vertex. 
A multivariate algorithm~\cite{BBDT,LHCb-PROC-2015-018} is used for the identification of secondary vertices consistent with the decay of a \bquark hadron.

Triggered data further undergo a centralised, offline processing step to deliver physics-analysis-ready data across the entire \lhcb physics programme~\cite{Stripping}.
In the offline selection, trigger signals are associated with reconstructed particles. 
Selection requirements can therefore be made on the trigger selection itself. Selection requirements can also be made on whether the decision is due to the signal candidate, referred to as `triggered on signal' (TOS), or if it is due to other particles produced in the $pp$ collision, referred to as `triggered independently of signal' (TIS). The decision can also be a combination of both (TIS\&TOS). 

Simulated events are used in the training of a multivariate selection algorithm, and for determining the shape of the kinematic and invariant-mass distributions of signal in $\Bdb\to \Dsm\Dp$ and $\Bsb\to \Dsp\Dm$ decays. In the simulation, $pp$ collisions are generated using \pythia~\cite{Sjostrand:2007gs,*Sjostrand:2006za} with a specific \lhcb configuration~\cite{LHCb-PROC-2010-056}. Decays of unstable particles are described by \evtgen~\cite{Lange:2001uf}, in which final-state radiation is generated using \photos~\cite{davidson2015photos}. The interaction of the generated particles with the detector, and its response, are implemented using the \geant toolkit~\cite{Allison:2006ve, *Agostinelli:2002hh} as described in Ref.~\cite{LHCb-PROC-2011-006}. The underlying \proton\proton interaction is reused multiple times, each with an independently generated signal decay~\cite{LHCb-DP-2018-004}.
The simulated $\BdorBsbar$ production kinematics are calibrated to match the observed spectrum of \mbox{$\Bdb\to\Dsm\Dp$} decays in background-subtracted data as a function of $(\pt,\eta)$ of the $b$ meson using a gradient boosted reweighter (GBR)~\cite{Rogozhnikov:2016bdp} technique.

\section{Candidate selection}

Charm mesons are obtained by combining three well-reconstructed tracks that form vertices detached from any PV.
Each of the tracks must be well identified by the particle identification (PID) system as either pions or kaons.
The reconstructed masses of the charm mesons are required to be within 15\mevcc of their known masses~\cite{PDG2024}, which coincides with around three times the corresponding mass resolution. 
Two charm meson candidates are then combined to reconstruct the decay vertex of the $B$ meson candidate, which is required to both be well separated from and point back to a PV, when combined with its momentum.  

Selected samples are composed of several sources of background in addition to the signals of interest. 
The combinatorial background consists of random combinations of tracks. This includes both candidates where either of the $D$ meson candidates is a spurious combination of tracks, and where both $D$ meson candidates are correctly reconstructed but do not originate from a $b$ meson. 
Other sources of background include the decays of $b$ mesons with additional photons that are not reconstructed, and must be included in the mass model. 

The combinatorial background is reduced using a multivariate selection based on a multilayer perceptron (MLP) classifier. 
The MLP is trained to identify signal decays using 15 variables, including properties of the $B$ meson, such as the significance of the displacement of the $B$ decay vertex from the PV, as well as of the final-state tracks, such as their transverse momenta.

The classifier is trained using simulated events as a proxy for the signal, and data in the upper-mass sideband of the $\BdorBsbar$ candidate ($ 5450 < m(\Dspm\Dmp) < 6000\mevcc$) as background. 
The MLP produces a discriminant for each candidate, with larger values indicating that a candidate is more consistent with the signal hypothesis. 
The optimal requirements on both this discriminant and the PID variables for each final-state track are determined by minimising the expected uncertainty on the $\Bsb$ asymmetry.
The uncertainty on the $\Bsb$ asymmetry is estimated as
\mbox{$N_{\Bsb}/\sqrt{N_{\Bsb} + N_{\text{Back}}}$}, 
where the yield of $\Bsb$ decays is estimated using the number of $\Bzb$ decays from a fit performed to the data and the known ratio of branching fractions, and $N_\text{Back}$ is the background yield within $3\sigma$ of the known $\Bsb$ mass \cite{PDG2024}, estimated using the upper-mass sideband of the selected $\Bsb$ candidates.
The selection requirement on the MLP retains around $80\%$ of the signal, while removing over $90\%$ of the background. 
Around $0.5\%$ of the selected candidates have a second candidate within the same event, typically a candidate composed of a subset of common particles with one or more exchanged with unrelated tracks in the event.
Both are retained for further analysis, as the fraction of such events is very low, and no significant bias towards either charge is observed. 

\section{Measurement of the raw asymmetries}
\label{sec_raw_asym}
The values of $A_{\text{raw}}(\Bdb\to \Dsm\Dp)$ and $A_{\text{raw}}(\Bsb\to \Dsp\Dm)$ are measured by performing a fit to the mass distributions of candidates in the range $5150 < m(\Dsmp\Dpm) < 5450\mevcc$. Components are included in the model for the \mbox{$\Bdb\to \Dsm\Dp$} and \mbox{$\Bsb\to \Dsp\Dm$} signal decays and their charge conjugates. A component is also introduced to model the partially reconstructed decay of $\Bd$ mesons to $\Dsspm \Dmp$ candidates, where the 
$\Dsspm$ decays further to a photon and a $\Dspm$ meson. 
Decays where heavier particles are not reconstructed, such as $\Bd\to\Dssp(\to\Dsp\piz)\Dm$, are found to have investigated and found to have insignificant yields for $m(\Dsmp\Dpm) > 5150\mevcc$ and thus are neglected. 
The final component models the combinatorial background.

Single-charm decays, such as $\Bzb \to \Dp\Km\Kp\pim$, are investigated using candidates that fail the requirements on the charm masses and flight distances. 
Few $\BdorBsbar$ candidates are observed, and extrapolating to the signal region, the expected yields are found to be negligible. 
Other processes, such as double-charm decays reconstructed by exchanging final-state particles between the two charm mesons, are investigated by recomputing the charm masses under all such hypotheses. 
No significant backgrounds of this type are observed.

The mass distributions for both of the signal components are described by the sum of a bifurcated Gaussian function with additional asymmetric tails and Gaussian function, as described in~\cite{LHCb-PAPER-2020-019}. 
Both functions share a common peak position, characterised by the most probable value (MPV) of the distribution. 
The right-width and the MPV of the $\Bd$ mass model are allowed to vary in the fit to data, while all other parameters are fixed relative to these parameters using simulation. 
Furthermore, the MPV of the $\Bsb$ signal is constrained with respect to that of the $\Bd$ using the known mass difference~\cite{PDG2024}. 

The combinatorial background is described by an exponential distribution, where the parameter of the exponential is determined in the fit to the data.
Partially reconstructed $\BdorBdbar \to \Dsspm ( \to \Dspm \gamma ) \Dmp$ decays, where the photon is not reconstructed, are present at low $m(\Dsmp\Dpm)$. 
Such decays are modelled by convolving a low-order polynomial function with the mass resolution distribution. 
The parameters of the polynomial function depend on the orbital structure and available phase space of the decay \cite{LHCb-PAPER-2017-021}. All parameters are determined from the simulation. 

\begin{figure}
    \centering
    \includegraphics[width=\textwidth]{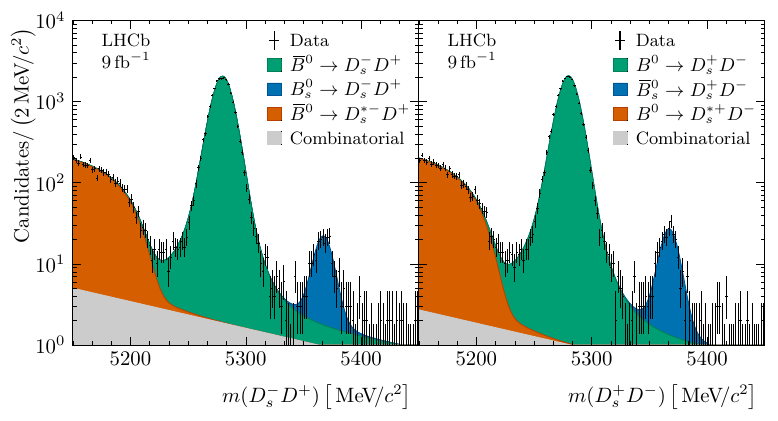}
    \caption{Mass distributions of the (left) $\Dsm\Dp$ and (right) $\Dsp\Dm$ systems. The fit model is also shown.}
    \label{fig:invariantmass}
\end{figure}

A binned maximum-likelihood fit to the $m(\Dsmp\Dpm)$ distributions is performed simultaneously to extract the time-integrated raw asymmetries of $\decay{\Bdb}{\Dsm\Dp}$ and $\decay{\Bsb}{\Dsp\Dm}$ decays. The invariant-mass distributions of the $\Dsp\Dm$ and  $\Dsm\Dp$ candidates and the corresponding shapes are shown in Fig.~\ref{fig:invariantmass}. The charge-integrated signal yields are $N(\Bdb\to \Dsm\Dp) = (43.0\pm 0.2)\times10^3$ and $N(\Bsb\to \Dsp\Dm) = (0.43\pm 0.02)\times 10^3$.

\section{Detection and production asymmetries}

The detection asymmetry is a consequence of the different efficiencies for particles and antiparticles traversing the full detector and subsequently being triggered, reconstructed and selected. 
Many sources of asymmetry approximately cancel due to the net uncharged final state, with the largest residual asymmetries arising from differences between kaon-pion pairs. 
The detection asymmetry, $A_\text{det}$ for the $\Km\pip$ pair can be decomposed into several sources 
\begin{equation}
 A_\text{det} = A_{K\pi} + A_\text{PID} + A_\text{trigger},
\end{equation}
where $A_{K\pi}$ is the tracking asymmetry of the kaon-pion pairs, $A_\text{PID}$ is the charge asymmetry associated with the efficiency of the PID requirements on the aforementioned kaon-pion pairs, and $A_\text{trigger}$ is the asymmetry in the efficiency of the trigger requirements between charges. 
All sources of asymmetry are assessed using calibration samples and are discussed below.

The interaction cross-section of kaons with matter varies rapidly at low $p_{\text{T}}$~\cite{PDG2024}, and the detection asymmetry for kaons is large. 
Resultantly, asymmetries arising from the hadronic interaction of final-state particles with the detector material, as well as track reconstruction and the detector acceptance, contribute to $A_{\PK\pi}$. 
The detection asymmetry of kaon-pion pairs is 
\begin{equation}
\label{eq_kpi_prec}
   A_{\PK\pi} \equiv \frac{\varepsilon(\Km \pip) - \varepsilon(\Kp \pim)}{\varepsilon(\Km \pip) + \varepsilon(\Kp \pim)}.
\end{equation}
This quantity is determined by measuring the difference in the values of $A_{\text{raw}}$ for the promptly produced $\Dp\to \Km \pip\pip$ and $\Dp\to \KS\pip$ decays, after weighting the two-dimensional distribution in $(\pt,\eta)$ of the $K\pi$ pair in the calibration sample to match the kinematics of the signal sample. 
This value is then corrected by the detection and \CP asymmetries associated with the $\KS$ meson ~\cite{LHCb-PAPER-2016-035}. 
The luminosity-weighted average of this asymmetry summed from all three $\PK\pi$ pairs in the final state is \mbox{$A_{\PK\pi} = - 0.0126 \pm 0.0019$}, where the uncertainty is due to the finite effective size of the calibration sample, following the kinematic weighting procedure. 

Since PID information is used in the candidate selection, there is a corresponding charge asymmetry ($A_{\text{PID}}$) associated with the efficiency of the PID selection criteria~\cite{LHCb-DP-2018-001}. 
This asymmetry is determined by measuring the efficiencies from a large sample of \mbox{$\Dstarp\to \Dz(\to \Km \pip)\pip$} decays applying the same PID requirements used in this analysis~\cite{LHCb-PUB-2016-021}, weighting the calibration sample to match the kinematics of the signal sample. 
The luminosity-weighted average of this asymmetry summed over all final-state tracks is \mbox{$A_\text{PID} = 0.0003 \pm 0.0011$}, where the uncertainty is determined by comparing the results from using two different binning schemes to weight the $\Dstarp\to \Dz(\to \Km \pip)\pip$ kinematic distributions.

The asymmetry associated with the hardware trigger, $A_\text{trigger}$ is further decomposed into 
\begin{equation}
 A_\text{trigger} = f_\text{TIS} A_\text{TIS} + f_\text{TOS} A_\text{TOS}, 
\end{equation}
where $f_\text{TIS}$ and $f_\text{TOS}$ are the fractions of candidates that would be recorded independent of the signal tracks, and candidates are only triggered by the signal, respectively, while $A_\text{TIS}$ and $A_\text{TOS}$ are the corresponding detection asymmetries. 
The detection asymmetry associated with subsequent stages of the trigger is negligible at the present precision. 

 The asymmetry associated with the TIS selections in the hardware trigger, $A_{\text{TIS}}$, is not dependent on the final state but could depend on the $\pt$ of the \PB candidate and the magnet polarity. 
The TIS asymmetry has been measured using $\Bm\to \Dzb\mun\neumb$ decays at $\sqs = 7$ and $8\tev$ for both magnet polarities in 20 bins of the $\pt$ of the $B$ meson, which is used to determine $A_\text{TIS}$ by integrating over the kinematics of the $B$ mesons. 
This yields \mbox{$A_{\text{TIS}} = -0.0002 \pm 0.0022$}.

The asymmetry associated with the TOS selections in the hardware trigger ($A_{\text{TOS}}$) is calculated from the TOS efficiencies in bins of $E_{\text{T}}$ measured in an unbiased sample of $\Dstarp\to \Dz(\to \Km \pip)\pip$ decays, for both charges of kaon and pion final states, and in each region of the HCAL~\cite{LHCb-DP-2018-001}. 
Here, the unbiasing is achieved by selecting only events that are recorded due to a muon trigger, independent of the decay of interest. 
The asymmetry of the TOS efficiency is calculated and summed for each final state kaon and pion over 20 \et bins in the range $0 < \et < 20\gev$. 
Measurements are made at $\sqs = 7, 8$ and $13\tev$ and the luminosity weighted average of this asymmetry is found to be  \mbox{$A_{\text{TOS}} = 0.0008 \pm 0.0027$}.
When combined with the TIS asymmetry, the total detection asymmetry associated with the hardware trigger is found to be $A_\text{trigger} = 0.0002 \pm 0.0017$.

As neutral mesons oscillate, the decay-time integrated, or effective, production asymmetry is the calculated accounting for the initial production asymmetry with the probability that the meson has oscillated into the flavour \cite{LHCb-PAPER-2011-029}. 
As the $\Bs$ oscillation period is significantly smaller than its lifetime ~\cite{PDG2024}, the effective production asymmetry is negligible.
Conversely, the effective production asymmetry of the $\Bdb$ is only mildly suppressed as the oscillation period of the $\Bd$ is comparable to its lifetime. 
The \Bzb production asymmetry at zero decay time has been measured at \mbox{$\sqs =7,8$ and $13\tev$} using \decay{\Bd}{\jpsi K^*(892)^0} decays~\cite{LHCb-PAPER-2016-062,LHCb-PAPER-2023-013}. 
The production asymmetries is provided in bins of rapidity and $\pt$ of the $\Bd$ meson for $7$ and $8$\tev, thus averaged over the kinematics in the signal sample, are found to be \mbox{$A_{\text{prod}}^{7\tev}(\Bzb) = -0.0016 \pm 0.0119 $} and \mbox{$A_{\text{prod}}^{8\tev}(\Bzb) = 0.0037 \pm 0.0074 $}.
The production asymmetry at 13\tev is found to be $A_\text{prod}^{13\tev}(\Bzb) = 0.0027 \pm 0.0013$.
The dilution of the production asymmetry due to $\Bz$ oscillations is estimated, accounting for the final-state specific decay-time acceptance and taking the $\Bd$ oscillation frequency from Ref.~\cite{PDG2024}. 
The dilution factor is found to be $\kappa = 0.449\pm0.003$, where the uncertainty is from the finite size of the simulated sample used to estimate the decay time acceptance. 
The luminosity-weighted effective production asymmetry is found to be $A_{\text{prod}}(\Bdb) = 0.0011 \pm 0.0007$, where the statistical and systematic uncertainties are combined. These measurements are compatible with other production asymmetry measurements~\cite{LHCb-PAPER-2023-040}.

\section{Systematic uncertainties and consistency checks}

\begin{table}
	\centering
	\caption{ \label{tab::cp_systematics}Absolute systematic uncertainties on the measured \CP asymmetries in $\Bdb\to\Dsm\Dp$ and $\Bsb\to \Dsp \Dm$ decays. The statistical uncertainties for each observable are included for comparison.}
	\begin{tabular}{ l@{}c@{}c }
		\toprule
		Source of uncertainty & \multicolumn{1}{c}{$\phantom{<\;}A_\CP(\Bdb\to \Dsm\Dp$)} & \multicolumn{1}{c}{$\phantom{<\;}A_\CP(\Bsb \to \Dsp \Dm$)} \\
		\midrule
        Signal model                            & $\phantom{<\;}0.0004$  & $\phantom{<\;}0.0090$ \\
        $\decay{\Bd} {\Dssp \Dm}$ model         & $<0.0001$ & $\phantom{<\;}0.0015$ \\
        Weighting of $\BdorBsbar$ kinematics    &  $\phantom{<\;}0.0011$  & $\phantom{<\;}0.0011$ \\
        $A_{K\pi}$                              & $\phantom{<\;}0.0019$  & $\phantom{<\;}0.0019$ \\
        $A_{\text{PID}}$                        & $\phantom{<\;}0.0026$  & $\phantom{<\;}0.0026$ \\
        $A_{\text{trigger}}$                    & $\phantom{<\;}0.0017$  & $\phantom{<\;}0.0017$ \\ 
        $A_{\text{prod}}$                       & $\phantom{<\;}0.0007$  & $<0.0001$ \\
        \midrule
        Total $\sigma_{\text{syst}}$            & $\phantom{<\;}0.0040$  & $\phantom{<\;}0.0099$ \\
        \midrule
        $\sigma_{\text{stat}}$                  & $\phantom{<\;}0.0053$  & $\phantom{<\;}0.0534$ \\
    \bottomrule
    \end{tabular}
\end{table}

The uncertainties associated with the choice of mass fit models are estimated by measuring the change in $A_{\text{raw}}$ when performing a fit with an alternative model, here a generalisation of the Crystal Ball function described in Ref.~\cite{Santos:2013gra}. In this alternative model, all parameters are constrained by simulation except the peak position, width and signal yields. The usage of the alternative model has a negligible impact on the measured asymmetries and thus no additional systematic uncertainties are applied. 

Uncertainties associated with the fit parameters of the signal shapes fixed from simulation are estimated by varying the parameters according to the covariance matrix from the fit to simulation, and then using these in a fit to data. 
This procedure is repeated many times, and the standard deviation of the resulting distribution is taken as the measure of the uncertainty. 
The extent to which the far larger $\Bz$ signal overlaps with the $\Bs$ signal is also affected by this uncertainty, and hence the impact on the determination of the $\Bs$ yield is far greater than on the $\Bz$ yield.  

The uncertainty from the shape of the $\Bd \rightarrow \Dss(\rightarrow \Dsp \gamma ) \Dm $ decay is evaluated by varying the resolution parameter of the shape by the uncertainty on the resolution of the signal shape from a fit to data. 
The systematic uncertainty is evaluated as the standard deviation of the resulting distributions when the $\Bd \rightarrow \Dss(\rightarrow \Dsp \gamma ) \Dm $ shape is varied 1000 times.

The systematic uncertainty from the weighting of the $\BdorBs$ production kinematics is evaluated by 
recomputing the weighting using only half of the data sample. 
This procedure is repeated 100 times, taking random halves of the data each time, and the standard deviation of the resulting ensemble is taken as the uncertainty on $A_{\CP}$.

Three sources of uncertainty on the detection asymmetry $A_{K\pi}$ are assessed. 
The uncertainty due to the size of the prompt $\mbox{\Dp\to \Km \pip\pip}$ and $\mbox{\Dp\to \KS\pip}$ samples is propagated from the invariant-mass fits used to calculate the raw asymmetries for the determination of $A_{\PK\pi}$. 
The uncertainty due to the simulation sample used to weight the prompt $\mbox{\Dp\to \Km \pip\pip}$ and $\mbox{\Dp\to \KS\pip}$ samples to match the kinematics of the signal in data is evaluated as the standard deviation in $A_{\PK\pi}$ when half of the simulation sample is randomly sampled 100 times. 
A systematic uncertainty is assigned to imperfections in the weighting of the prompt $\Dp\to \Km \pip\pip$ and $\Dp\to \KS\pip$ kinematic variables. This is assessed as the change in $A_{\PK\pi}$ when the prompt $\Dp\to \Km \pip\pip$ and $\Dp\to \KS\pip$ samples are weighted using binned methods instead of the GBR technique. 

The systematic uncertainty associated with the detection asymmetry from the hardware trigger is the combination of the uncertainties on the different trigger categories, weighted by the fraction of candidates in each category. 
The dominant sources of systematic uncertainty are from the external inputs, which are the binned asymmetries for $\mbox{\Bm\to \Dzb\mun\neumb}$ decays and for the flavour-specific charge asymmetries in the HCAL for TIS and TOS, respectively. 
A small additional contribution to the uncertainty comes from the finite size of the simulated sample used to calculate these asymmetries for the kinematics of the signal decay. 

The uncertainty from the finite size of the $\Dstarp\to \Dz(\to \Km \pip)\pip$ sample used to calculate $A_{\text{PID}}$ is propagated onto the detection asymmetries.
The uncertainty associated with the size of the simulation sample used to weight the kinematics of the $\mbox{\Dstarp\to \Dz(\to \Km \pip)\pip}$ sample is evaluated as the standard deviation in $A_{\text{PID}}$ when half of the simulation sample is randomly sampled 100 times.
The PID efficiencies of each final-state track cannot be assumed to be completely uncorrelated. The bias on the $A_{\CP}$ measurement associated with this correlation estimated as the change in $A_{\text{PID}}$ when applying all of the track PID requirements at the same time, versus the sum of the individual track $A_{\text{PID}}$ when applying one track PID requirement at a time. The statistical uncertainty on the bias is calculated by measuring the standard deviation of the difference between the two PID asymmetries when half of the simulation sample is randomly sampled 100 times. 
The different sources of uncertainty on the PID asymmetries are added in quadrature and tabulated in Table~\ref{tab::cp_systematics}, where the dominant uncertainty is due to the finite size of the calibration sample.

The systematic uncertainties on $A_{\text{prod}}$ are obtained from the measurements of Ref.~\cite{LHCb-PAPER-2023-040} and the uncertainty on the dilution factor from finite simulated statistics. 
The uncertainty from the $\Bz$ oscillation frequency has a negligible impact on the dilution factor. 
The uncertainties are propagated onto $\mbox A_{CP}(\Bdb\rightarrow\Dsm\Dp) $, while the impact on \mbox{$A_{CP}(\Bsb\rightarrow\Dsp\Dm)$} is estimated by repeating the fit many times, varying the $\Bzb$ production asymmetry, and are found to be negligible. 

The uncertainties on the integrated luminosity per year and magnet polarity, which are around $0.04\invfb$, are propagated to the detection and production asymmetries, and are found to be negligible compared to the total uncertainty.  
The full list of systematic and statistical uncertainties is given in Table~\ref{tab::cp_systematics}.

The analysis is repeated, dividing the data into statistically independent subsets, such as the magnet polarity and centre-of-mass energy, trigger category, and $B$ kinematics. 
All corrections are recomputed for each subset, and the compatibility in the determination of the asymmetries in each subset is calculated. 
Ten such divisions are performed, with $p$-values for the consistency between subsets varying between 5 and 80\%, indicating good consistency in the determination of parameters and the procedures to correct for nuisance asymmetries.

\section{Results and conclusions}

In summary, using a $pp$ collision data sample corresponding to an integrated luminosity of $9\invfb$ at $\sqrt{s} = 7,8$ and $13\tev$ collected by the LHCb experiment, the \CP asymmetries of $\Bdb\to \Dsm\Dp$ and $\Bsb\to\Dsp\Dm$ have been measured to be
\begin{align*}
\begin{split}
A_{\CP}(\Bdb\to \Dsm\Dp) &= 0.0009 \pm 0.0053 \pm 0.0040, \\
A_{\CP}(\Bsb\rightarrow \Dsp \Dm ) &= 0.103\phantom{0} \pm 0.053\phantom{0} \pm 0.010,
\end{split}
\end{align*}
where the first uncertainty is statistical, and the second is systematic. The correlation between the measurements is $-0.0023$. Both measurements are consistent with both \CP symmetry and the predictions made in Ref.~\cite{davies2023bar}. 
The measurement of \mbox{$A_{\CP}(\Bdb\to \Dsp \Dm )$} is more precise than the current best determination by more than a factor of three~\cite{rohrken2012measurements}, while differing in central value by less than one standard deviation. 
The measurement of \mbox{$A_{\CP}(\Bsb\to \Dsp \Dm )$} is the first performed. 
These measurements significantly improve knowledge of $B \to D \Dbar$ decays, which can help constrain BSM models~\cite{Fu_2011, Li_2010, Lu_2011, PhysRevD.79.055004, xu2016studying, davies2023bar}.

\section*{Acknowledgements}
%
%
\noindent We express our gratitude to our colleagues in the CERN
accelerator departments for the excellent performance of the LHC. We
thank the technical and administrative staff at the LHCb
institutes.
We acknowledge support from CERN and from the national agencies:
ARC (Australia);
CAPES, CNPq, FAPERJ and FINEP (Brazil); 
MOST and NSFC (China); 
CNRS/IN2P3 and CEA (France);  
BMFTR, DFG and MPG (Germany);
INFN (Italy); 
NWO (Netherlands); 
MNiSW and NCN (Poland); 
MEC/IFA (Romania); 
MICIU and AEI (Spain);
SNSF and SER (Switzerland); 
NASU (Ukraine); 
STFC (United Kingdom); 
DOE NP and NSF (USA).
We acknowledge the computing resources that are provided by ARDC (Australia), 
CBPF (Brazil),
CERN, 
IHEP and LZU (China),
IN2P3 (France), 
KIT and DESY (Germany), 
INFN (Italy), 
SURF (Netherlands),
Polish WLCG (Poland),
IFIN-HH (Romania), 
PIC (Spain), CSCS (Switzerland), 
GridPP (United Kingdom),
and NSF (USA).  
We are indebted to the communities behind the multiple open-source
software packages on which we depend.
Individual groups or members have received support from
RTP (Australia), 
Key Research Program of Frontier Sciences of CAS, CAS PIFI, CAS CCEPP (China); 
Minciencias (Colombia);
EPLANET, Marie Sk\l{}odowska-Curie Actions, ERC and NextGenerationEU (European Union);
A*MIDEX, ANR, IPhU and Labex P2IO, and R\'{e}gion Auvergne-Rh\^{o}ne-Alpes (France);
Alexander-von-Humboldt Foundation (Germany);
ICSC (Italy); 
Severo Ochoa and Mar\'ia de Maeztu Units of Excellence, GVA, XuntaGal, GENCAT, InTalent-Inditex and Prog.~Atracci\'on Talento CM (Spain);
the Leverhulme Trust, the Royal Society and UKRI (United Kingdom).

\addcontentsline{toc}{section}{References}
\bibliographystyle{LHCb}
\bibliography{main,standard,LHCb-PAPER,LHCb-CONF,LHCb-DP,LHCb-TDR}

\newpage
\centerline
{\large\bf LHCb collaboration}
\begin
{flushleft}
\small
R.~Aaij$^{38}$\lhcborcid{0000-0003-0533-1952},
A.S.W.~Abdelmotteleb$^{57}$\lhcborcid{0000-0001-7905-0542},
C.~Abellan~Beteta$^{51}$\lhcborcid{0009-0009-0869-6798},
F.~Abudin\'en$^{57}$\lhcborcid{0000-0002-6737-3528},
T.~Ackernley$^{61}$\lhcborcid{0000-0002-5951-3498},
A. A. ~Adefisoye$^{69}$\lhcborcid{0000-0003-2448-1550},
B.~Adeva$^{47}$\lhcborcid{0000-0001-9756-3712},
M.~Adinolfi$^{55}$\lhcborcid{0000-0002-1326-1264},
P.~Adlarson$^{85}$\lhcborcid{0000-0001-6280-3851},
C.~Agapopoulou$^{14}$\lhcborcid{0000-0002-2368-0147}
% ...

{\footnotesize \it

% ...
$^{14}$Universit{\'e} Paris-Saclay, CNRS/IN2P3, IJCLab, Orsay, France\\
% ...
$^{38}$Nikhef National Institute for Subatomic Physics, Amsterdam, Netherlands\\
% ...
$^{47}$Instituto Galego de F{\'\i}sica de Altas Enerx{\'\i}as (IGFAE), Universidade de Santiago de Compostela, Santiago de Compostela, Spain\\
% ...
$^{51}$Physik-Institut, Universit{\"a}t Z{\"u}rich, Z{\"u}rich, Switzerland\\
% ...
$^{55}$H.H. Wills Physics Laboratory, University of Bristol, Bristol, United Kingdom\\
% ...
$^{57}$Department of Physics, University of Warwick, Coventry, United Kingdom\\
% ...
$^{60}$School of Physics and Astronomy, University of Glasgow, Glasgow, United Kingdom\\
$^{61}$Oliver Lodge Laboratory, University of Liverpool, Liverpool, United Kingdom\\
% ...
$^{69}$Syracuse University, Syracuse, NY, United States\\
% ...
$^{85}$Department of Physics and Astronomy, Uppsala University, Uppsala, Sweden, associated to $^{60}$\\
% ...
}
\end{flushleft}

\end{document}